\RequirePackage[mathlines]{lineno} 
\documentclass[12pt]{iopart}

\usepackage{graphicx}
\begin{document}

\title[PHENIX Measurements of Higher-order Flow Harmonics]{PHENIX Measurements of Higher-order Flow Harmonics in Au+Au collisions at  $\mathbf{\sqrt{s_{NN}} = 200}$~GeV}
\author{Roy Lacey for the PHENIX Collaboration}
\address{Chemistry Dept,
Stony Brook University,
Stony Brook, NY 11794-3400,
USA}
\ead{Roy.Lacey@Stonybrook.edu}

\begin{abstract}

Flow coefficients $v_n$ for $n$ =~2,~3,~4, characterizing the
anisotropic collective flow in Au+Au collisions at $\sqrt{s_{NN}}~=~200$ GeV, 
are presented. They indicate the expected growth 
of viscous damping for sound propagation in the quark gluon plasma (QGP) produced 
in these collisions. Hydrodynamical model comparisons which include the effects 
of initial state geometry fluctuations, highlight the 
role of higher harmonics ($v_{n, n>2}$) as a constraint for disentangling the effects of viscosity and 
initial conditions, and suggest a small specific viscosity for the QGP. 
This viscosity is compatible with that obtained via a newly proposed technique \cite{Lacey:2011ug} which 
employs the relative magnitudes of $v_n$ to estimate the viscosity, and the ``viscous horizon'' 
or length-scale which characterizes the highest harmonic that survives viscous damping. 

\end{abstract}


\vspace{-1 cm}

\section{Introduction}

Measurements of anisotropic flow in heavy-ion collisions at the Relativistic Heavy Ion collider (RHIC), 
continue to play a central role in ongoing efforts to characterize the transport properties of the 
quark gluon plasma (QGP) produced in these collisions. Recently, considerable attention has been 
given to extractions of the specific viscosity $\eta/s$ [the ratio of viscosity ($\eta$) 
to entropy density ($s$)] via elliptic flow ($v_2$) measurements. These extractions indicate 
a range of 1-2 times the conjectured lower bound \cite{Kovtun:2004de} for $\eta/s$, 
({\em i.e.} $4\pi\frac{\eta}{s}  \sim 1-2$). This large uncertainty (100\%) is known to be dominated 
by the uncertainty in model estimates of the initial eccentricity \cite{Song:2010mg,Lacey:2010fe}.
Thus, a more precise extraction of $\eta/s$ requires the development of new experimental constraints.

	Because of the acoustic nature of flow ({\em i.e.} it is driven by pressure gradients),
a transparent way to evaluate the strength of dissipative effects, is to consider the 
attenuation of sub-horizon sound modes in the plasma. In the presence of viscosity ($\eta$), 
sound intensity is exponentially damped as :
$
\delta T_{\mu\nu} (t) = \exp{\left(-{2 \over 3} {\eta \over s} {k^2 t \over  3T } \right)}  \delta T_{\mu\nu} (0)
$
\cite{Staig:2010pn}, where the spectrum of initial (t = 0) perturbations of the energy-momentum tensor $T_{\mu\nu}$, 
can be associated with the harmonics of the shape deformations and density fluctuations of the 
collision zone; $k$ is the wave number for these harmonics, and $t$ and $T$ are the expansion 
time and the temperature of the plasma respectively. Since viscous damping scales as $k^2$ 
the viscous corrections for the eccentricity driven harmonics $v_n$
(with wavelengths $2\pi {\bar R}/n$ for $n\ge 1$, {\em i.e.} $k \sim n/{\bar R}$), 
are expected to scale as $n^2K$; ${\bar R}$ is the transverse  size of the 
collision zone and $K$ is the Knudsen number \cite{Lacey:2011ug}. The latter 
is often used to parametrize viscous corrections.
The length scale $r_v$ or {\em ``viscous horizon''} separates the sound 
wavelengths which are effectively damped out, from those which are 
not, and $k_v = 2\pi/r_v$ is linked to the order ($n_v$) 
of the highest harmonic which survives viscous damping. Thus, the relative 
magnitudes of the higher-order harmonics ($v_{n, n\ge 3}$) are expected to provide 
additional constraints on both the magnitude of $\eta/s$ and the ``best'' 
model for eccentricity determinations \cite{Staig:2010pn,Lacey:2010hw,Alver:2010dn}.
Here, we report new $v_n$ measurements \cite{Adare:2011tg} and investigate their utility as 
constraints for precision extraction of $\eta/s$. 

\vspace{-0.3 cm}
%
\section{Data analysis and results}

The results presented are derived from $\sim 3.0 \times 10^9$  Au+Au events
obtained during the 2007 RHIC running period. The Fourier coefficients $v_n$ were 
obtained via (i) pair-wise distributions in the azimuthal angle difference 
($\Delta\phi =\phi_a - \phi_b$) between particles with pseudorapidity 
separation $\Delta\eta' > 1$ and transverse momenta $p^a_{T}$ and 
$p^b_{T}$ (respectively);
%
$
\frac{dN^{pairs}}{d\Delta\phi} \propto 
\left( 1 + \sum_{n=1}2v^a_n(p^a_T)v^b_n(p^b_T)\cos(n\Delta\phi) \right),
%
$
and (ii) azimuthal distributions 
$
\frac{dN}{d\phi} \propto \left( 1 + \sum_{n=1}2v_n\cos(n\phi - n\Psi_n) \right),
%
$
of charged hadrons detected in the 
PHENIX central arms at an azimuthal angle $\phi$, relative to event 
planes $\Psi_{n}$ \cite{Ollitrault:1992bk} obtained with three separate 
detectors: Beam-Beam Counters (BBC), Reaction-Plane Detectors (RXN), 
and Muon Piston Calorimeters (MPC). 
Each detector has a North (South) component which allows correlation studies 
between sub-event planes $\Phi_n$ determined at forward (backward) rapidity. 
The absolute pseudorapidity coverage for these detectors are $3.1 < \left| \eta'_{_{\rm BBC}}
\right| < 3.9 $, $1.0 < \left| \eta'_{{\rm RXN}} \right| < 2.8$ and $3.1
< \left| \eta'_{_{\rm MPC}} \right| < 3.7 $.
%
%
%
\begin{figure}[h!]
\begin{center}
\includegraphics[width=0.5\linewidth]{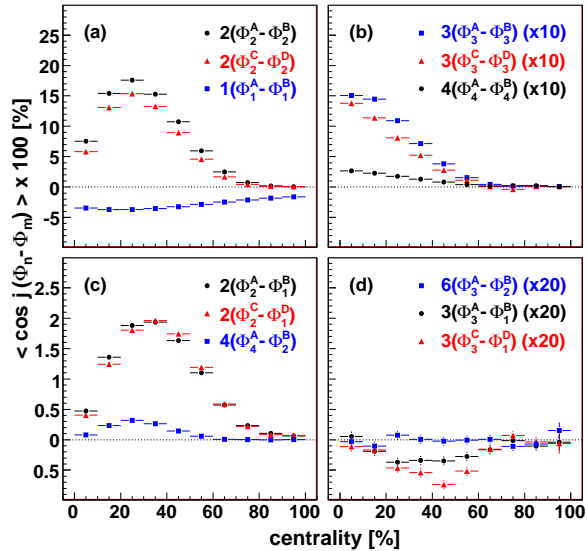}
\caption{Correlation strengths of the event planes for various detector combinations as 
a function of collision centrality. The detectors in which the event plane is 
measured are: RXN North (A), BBC South (B), MPC North (C) and MPC South (D). 
}
\label{fig1}
\end{center} 
\end{figure}
\vspace{-0.3 cm}

Figure \ref{fig1} shows the centrality dependence of the 
correlation strength $\left\langle \cos(j(\Phi_n^{A}-\Phi_m^{B}))\right\rangle$ 
for sub-event combinations ($A,B$) involving different event-plane detectors 
with $\Delta\eta' \sim 5$ and  $\Delta\eta' \sim 7$. 
Note that the order $j$ is chosen to account for the $n$-multiplet of 
directions ($2\pi/n$) of $\Phi_n$ and the magnitudes for 
$\left\langle \sin(j(\Phi_n^{A} -\Phi_m^{B}))\right\rangle$ are 
consistent with zero for all centrality, $j$, and $\Phi$
combinations. Positive sub-event correlations are indicated
in panels (a) and (b) for $\Psi_{2,3,4}$, with sizable magnitudes
for $\Psi_{2,3}$ and much smaller values for $\Psi_4$. 
The negative correlation indicated in panel (a) is due
to the well known antisymmetric pseudorapidity dependence of
sidewards flow ($v_1$), as well as momentum conservation.
The expected correlation between $\Psi_1$ and
$\Psi_2$, and that between $\Psi_2$ and $\Psi_4$ are 
confirmed in panel (c); they show that $\Psi_1$, $\Psi_2$ and $\Psi_4$ are 
correlated with the reaction plane. An initial state fluctuation origin 
of $\Psi_3$ [and hence $v_3$] is well supported by the absence of 
a correlation between $\Psi_2$ and $\Psi_3$ in panel (d). The absence of a correlation 
between $\Psi_2$ and $\Psi_3$ reflects the rather large fluctuations of $\Psi_3$ about $\Psi_2$ 
and gives a null value for $v_3$ measured relative to $\Psi_2$ \cite{Lacey:2010av}.

\vspace{0.3 cm}
\begin{figure}[h!]
\begin{center}
\includegraphics[width=0.8\linewidth]{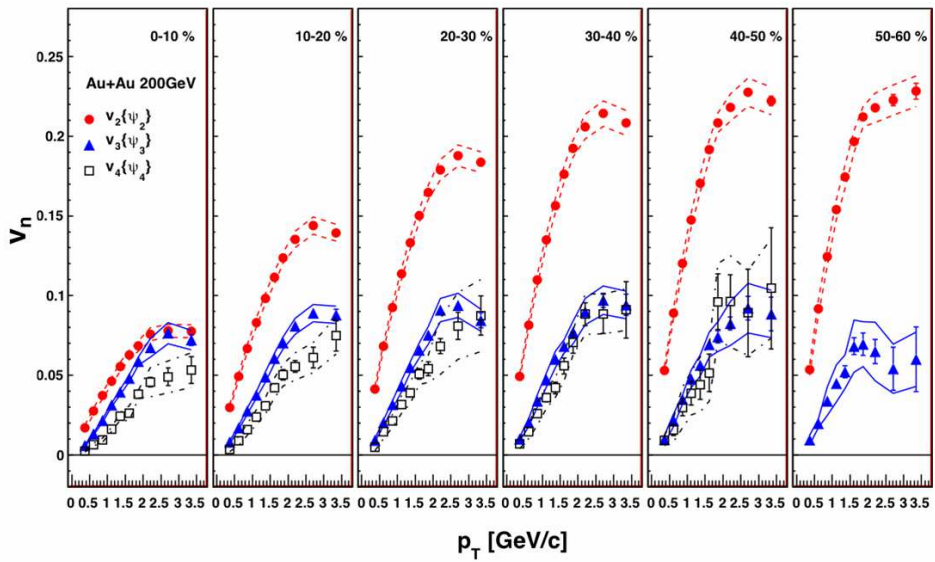}
\caption{$v_n$ vs. $p_T$ for several centrality bins as indicated. 
}
\label{fig2}
\end{center} 
\end{figure}
\vspace{-0.3 cm}

\begin{figure}[h!]
\includegraphics[height=6.9cm,width=8cm]{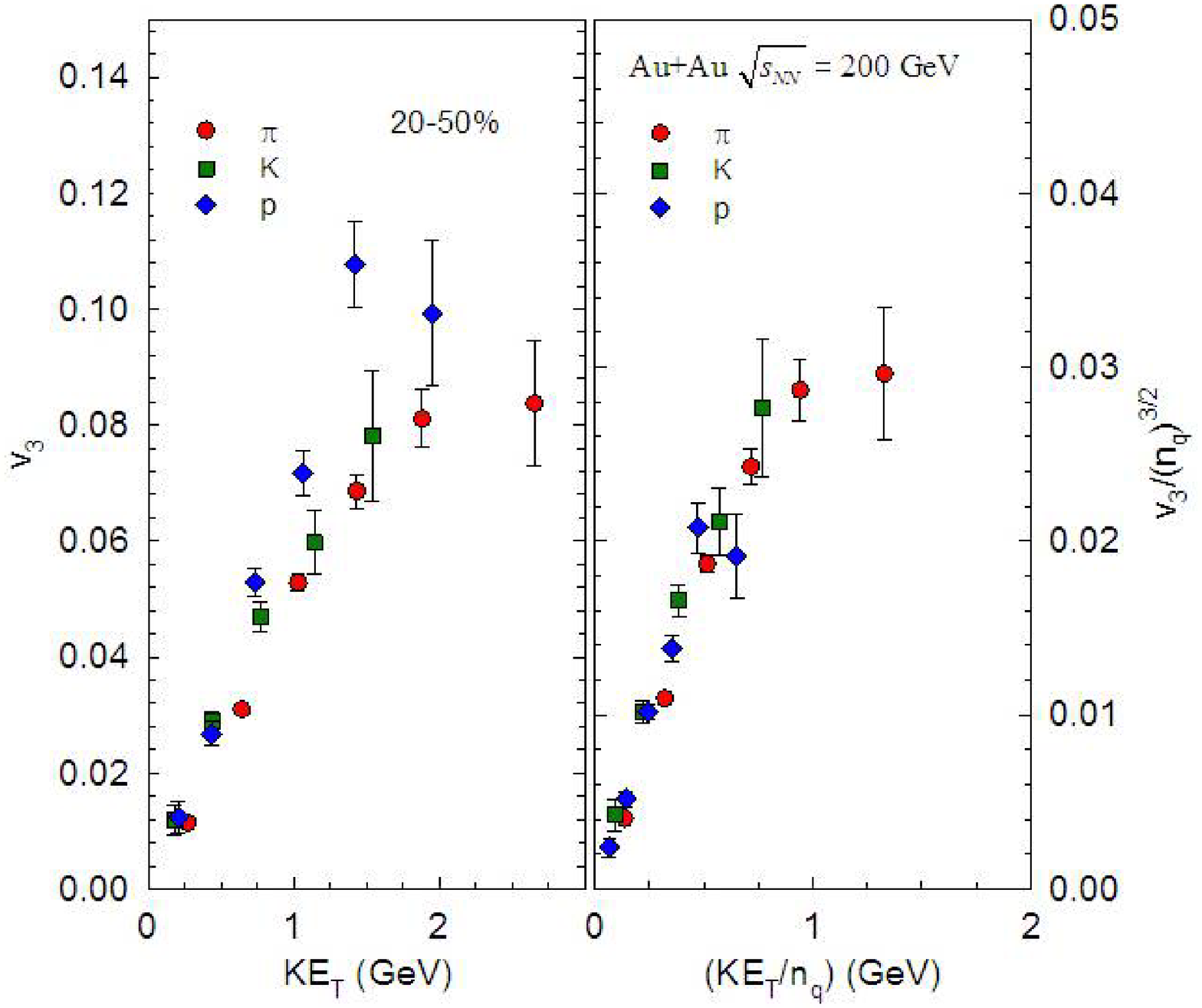}
\includegraphics[height=7.2cm,width=8cm]{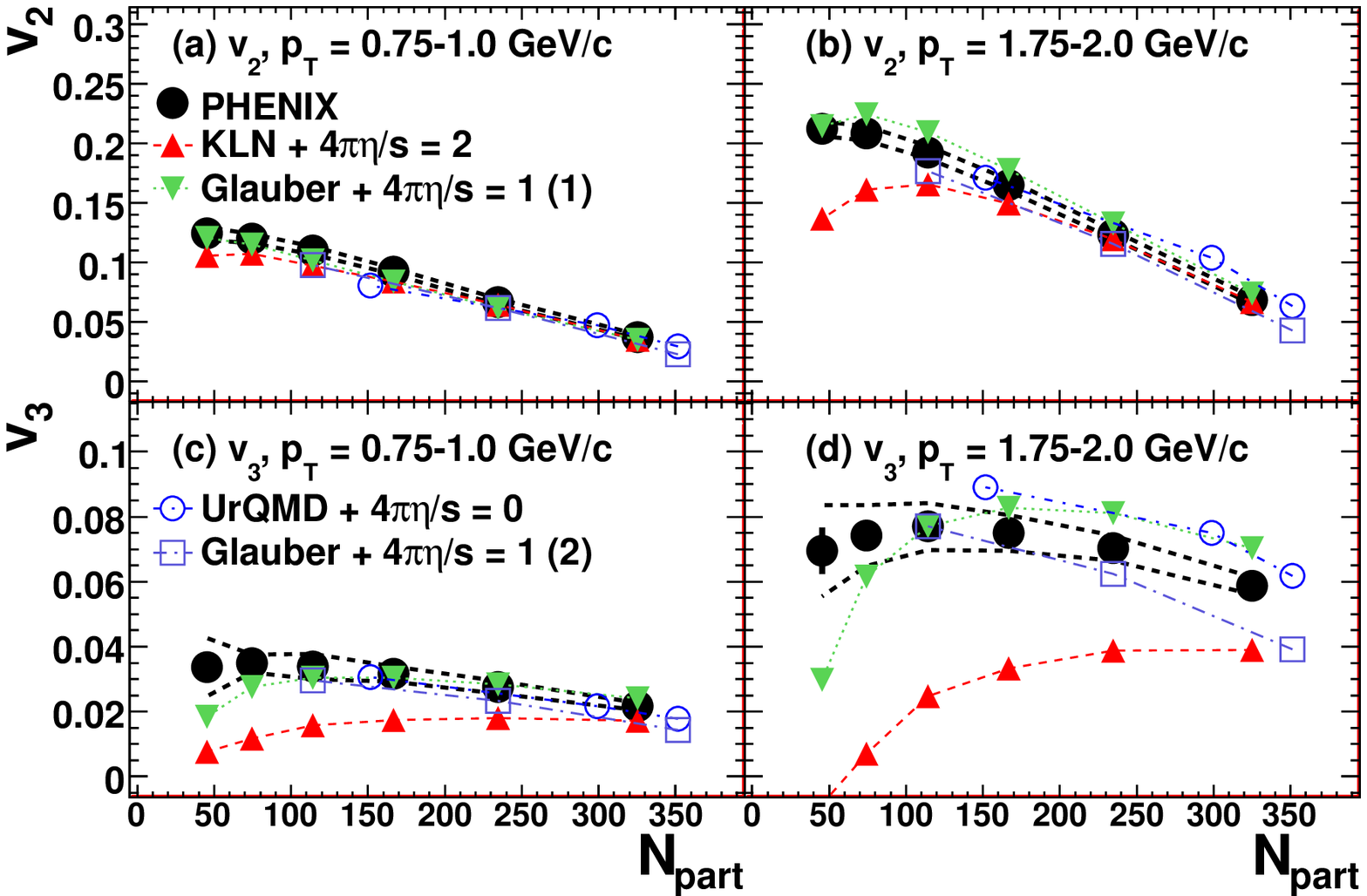}
\caption{Illustration of quark number scaling for $v_3$ (left). 
Comparisons of $v_{n}\{\Psi_n\}$  vs.  $N_{part}$ with theory as indicated (right). 
}
\label{fig3} 
\end{figure}

	Figure \ref{fig2} shows a clear growth of $v_2$ from central to mid-peripheral collisions, and a near
constancy of $v_3$ and $v_4$ which are strong indicators of the role of initial
state fluctuations in establishing the higher harmonics. 
It is noteworthy that 
the ratios $v_3/(v_2)^{3/2}$ and $v_4/(v_2)^2$ are essentially independent of 
$p_T$ (for $p_T < 2.5-3.0$ GeV/c) but do increase rapidly from peripheral to 
central collisions. These scaling patterns have been interpreted as an indication that viscous 
damping for the higher harmonics follow the expected acoustic (or $n^2K$) 
scaling \cite{Lacey:2011ug}. Interestingly, these scaling patterns are also reflected 
in quark number ($n_q$) scaling of particle identified data. That is, $v_n/(n_q)^{n/2}$ for different 
particle species, plotted as a function of transverse kinetic energy $KE_T$, gives 
essentially a single curve as illustrated for $v_3/(n_q)^{3/2}$ in the left panel of Fig. \ref{fig3}. 
The centrality dependence of $v_3/(v_2)^{3/2}$ and $v_4/(v_2)^2$ also provide a valuable 
constraint for eccentricity models \cite{Lacey:2011ug}; they favor a Glauber initial 
eccentricity model.

%
The right panels of Fig.~\ref{fig3} compare the centrality dependence of
$v_{2}\{\Psi_2\}$ and $v_{3}\{\Psi_3\}$ with several hydrodynamical model 
calculations \cite{Schenke:2010rr,Alver:2010dn}, which demonstrates an essential 
constraint that the data provide.
Panel (a) shows good agreement between data and calculations which employ MC-Glauber and 
MC-KLN initial eccentricities ($\varepsilon_2$) paired with viscosity values of
$4\pi\frac{\eta}{s}$ = 1 and 2, respectively. The resulting $\eta/s$ uncertainty ($\sim 100\%$)
reflect the model dependence of $\varepsilon_2$.
Differences between the calculations become more apparent for the higher $p_T$ 
selection shown in panel (b), but the implied uncertainty for $\eta/s$ remains 
large. Panels (c) and (d) demonstrates the utility of $v_{3}\{\Psi_3\}$ as a 
constraint for disentangling the effects of viscosity and initial conditions; 
they indicate excellent agreement with the results from viscous
hydrodynamics which employ Glauber initial eccentricities and $4\pi\frac{\eta}{s} = 1$,
and rather poor agreement with calculations which employ MC-KLN initial conditions 
and $4\pi\frac{\eta}{s}= 2$. The constraining power of $v_{3}\{\Psi_3\}$ stems from 
the fact that viscous corrections to $v_n$ scale as $n^2$ and the two eccentricity 
models give similar values for $\varepsilon_3$ \cite{Lacey:2010hw}. Thus, the larger 
viscosity needed for agreement with the data with MC-KLN eccentricities in panels (a) and (b), 
leads to a significant under prediction of $v_{3}\{\Psi_3\}$ in panels (c) and (d).
In summary, PHENIX $v_n$ measurements provide important constraints
for robust extraction of $\eta/s$.

\vspace{-0.3 cm}
%

\end{document}